\documentclass[aps,twocolumn,prl,amssymb,showpacs]{revtex4-1}
\usepackage{graphpap}
\usepackage[dvips]{graphicx}
\usepackage[dvips]{graphics}
\usepackage{color}
\usepackage{amsmath}
\usepackage[normalem]{ulem}

\newcommand{\newer}[1]{ {#1} }

\begin{document}

\title{Limitations to Carrier Mobility and Phase-Coherent Transport in Bilayer Graphene}
\author{S. Engels$^{1,2}$, B. Terr\'es$^{1,2}$, A. Epping$^{1,2}$, T. Khodkov$^{1,2}$, K.~Watanabe$^3$, T. Taniguchi$^3$, B. Beschoten$^{1}$ and C. Stampfer$^{1,2}$}
\affiliation{
$^1$JARA-FIT and 2nd Institute of Physics, RWTH Aachen University, 52074 Aachen, Germany, EU \\
$^2$Peter Gr\"unberg Institute (PGI-9), Forschungszentrum J\"ulich, 52425 J\"ulich, Germany, EU \\
$^3$National Institute for Materials Science, 1-1 Namiki, Tsukuba, 305-0044, Japan 
}

\begin{abstract}
We present transport measurements on high-mobility bilayer graphene fully encapsulated in hexagonal boron nitride. We show two terminal quantum Hall effect measurements which exhibit full symmetry broken Landau levels at low magnetic fields. From weak localization measurements, we extract gate-tunable phase coherence times $\tau_{\phi}$ as well as the inter- and intra-valley scattering times $\tau_i$ and $\tau_*$. While $\tau_{\phi}$ is in qualitative agreement with an electron-electron interaction mediated dephasing mechanism, electron spin-flip scattering processes are limiting $\tau_{\phi}$ at low temperatures. The analysis of $\tau_i$ and $\tau_*$ points to local strain fluctuation as the most probable mechanism for limiting the mobility in high-quality bilayer graphene.
\end{abstract}

 \maketitle

\newpage

\begin{figure}[h]\centering
\includegraphics[draft=false,keepaspectratio=true,clip,
         width=1\linewidth]%
                   {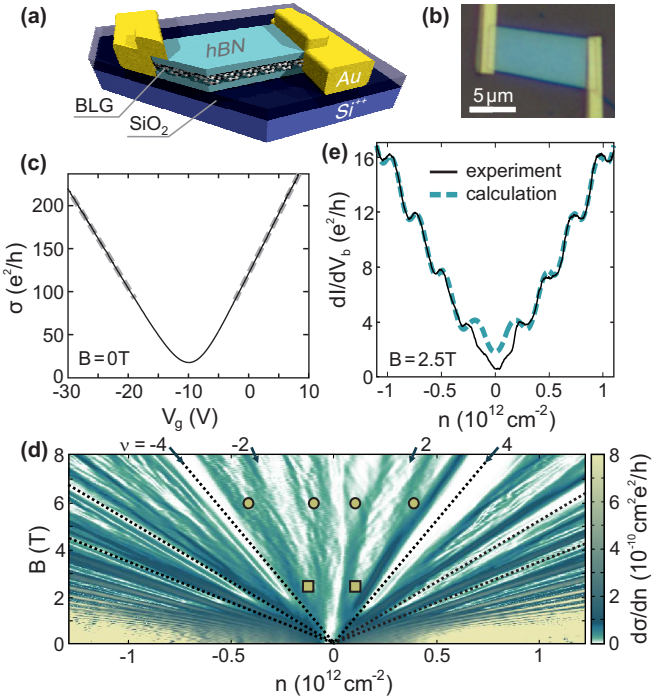}
\caption[fig01]{(color online) (a) Illustration and (b) optical image of a contacted hBN-BLG-hBN heterostructure.
(c) Conductivity $\sigma$ vs. $V_{g}$ of the device shown in (b) at $T$~=~77~K. Here we subtracted a contact resistance $R_C = 600~\rm{\Omega}$ for electrons and 650~$\rm{\Omega}$ for holes by following Ref.~\cite{cas10}. For more details on $R_C$ please see supplementary information. (d) Transconductivity $d\sigma/dn$ as function of $n$ and perpendicular $B$-field. (e) Solid line shows $dI/dV_{b}$ in dependence of $n$ at $B$~=~2.5~$T$. The result of a theoretical calculation is shown by the dashed line. 
}
\label{fig01}
\end{figure}

\begin{figure}[tb]\centering
\includegraphics[draft=false,keepaspectratio=true,clip,%
                   width=1\linewidth]%
                   {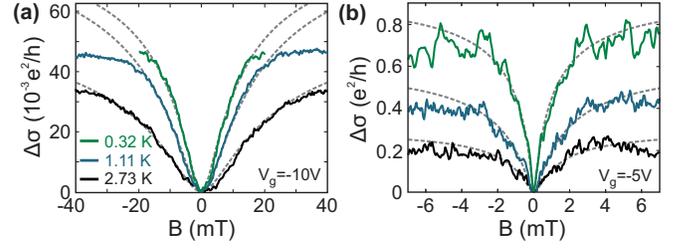}
\caption[fig02]{(color online) Magneto-conductivity at low magnetic fields and different temperatures $T=0.3$~K, $1.1$~K and $2.7$~K for (a) $V_g = -10$~V (charge carrier density $n$ close the charge neutrality point) and (b) $V_g = -5$~V (corresponding to $n=0.32 \times \rm{10}^{12} \rm{cm}^{-2}$). Fits to the weak localization model for BLG are displayed by the dotted lines.}
\label{fig02}
\end{figure}

Bilayer graphene (BLG) is an interesting material system to explore phase-coherent mesoscopic transport with unique electronic properties~\cite{mcc13}.
In contrast to single-layer graphene, in BLG a band gap can be opened by an external electric
field~\cite{cas09,oos07} making local depletion of the two-dimensional electron gas (2DEG)
possible similar to III/V heterostructures. This is an important prerequisite for implementing state-of-the-art phase-coherent quantum device concepts~\cite{ned07,ji03}. 
In contrast to conventional 2DEGs, the massive Dirac fermion nature of the quasi-particles in BLG 
results in an unconventional
quantum Hall effect~\cite{nov06,mcc06} and promises unique quantum interference properties~\cite{nan11}.
So far, the observable transport phenomena in BLG devices suffer from the limited device quality which is most likely a consequence of the high sensitivity of BLG on the surrounding environment.
Recent developments in device fabrication have shown that a significant improvement in sample quality can be obtained by replacing conventional SiO$_{2}$ with hexagonal boron nitride (hBN)~\cite{dea10}. This material provides an ultra-flat
substrate for graphene~\cite{dea10,xue11} and enables the realization of the high-mobility samples that are required to study, e.g. quantum phase transitions in the lowest 
quantum Hall state~\cite{mah13} 
or superlattice effects such as the Hofstadter butterfly~\cite{dea13,hun13,pon13}. However,
despite these improvements, it remains 
difficult to experimentally address the 
microscopic mechanisms that limits carrier mobility 
and  phase-coherence in high-quality BLG.\\
To address these important questions, we present diffusive transport measurements on BLG fully encapsulated in hBN. Our fabrication technique allows to obtain high-mobility samples, which show a well-developed quantum Hall effect and a full degeneracy breaking of the zero Landau level around $B$~=~6~T. To investigate the limits of phase-coherent transport and to gain insights on the limitations to carrier mobility in these 
devices, we perform weak localization measurements~\cite{gor07}. 
From these measurements, we extract the inter- and intra-valley scattering times, as well as the phase-coherence time. Our results indicate (i) that the main sources of dephasing in high-quality BLG are the electron-electron interaction as well as electron spin-flip scattering, and (ii) that mobility is not limited by inter-valley scattering processes. Moreover, we observe that the mean-free path quantitatively matches the intra-valley scattering length over a wide range of carrier densities. Our findings point at intra-valley scattering  as the main limitation to mobility in BLG. We discuss local strain fluctuations as the possible source of these mobility-limiting scattering events. \newline
\newer{ Here, we present two-terminal measurements on bilayer graphene encapsulated in hBN, see schematic in Fig.~\ref{fig01}(a). 
The investigated device has length $L~\approx~16~\rm{\mu m}$ and width $W~\approx~7~\rm{\mu m}$, see Fig. 1(b).
Detailed information of the sample fabrication, which is similar to Ref.~\cite{wan13}, are given in the supplementary information. All measurements are performed in a $\rm{He^{3}}$ cryostat with a base temperature of $T = 300$~mK (unless stated otherwise).}
In Fig.~\ref{fig01}(c) we show the conductivity of this device as function of back-gate voltage $V_{g}$ at $T$~=~77~K. 
The charge neutrality point is at a back-gate voltage $V_{g}^0 = -10$~V, indicating electron doping in our sample. 
By taking a gate lever arm
of $\alpha= 6.5 \times 10^{10}$ $\rm{cm}^{-2}$V$^{-1}$ (see details below) we extract
from the linear increase of the conductivity above $|\Delta V_g| = |V_g-V_{g}^0| \approx 10$~V a lower limit of the hole mobility of $\mu_h \approx 40,000$~$\rm{cm}^2/\rm{Vs}$ and of the electron mobility  of $\mu_e \approx 50,000$~ $\rm{cm}^2/\rm{Vs}$ (see dashed lines in Fig.~\ref{fig01}(c)). 
These mobility values have been reproduced on a number of encapsulated BLG devices, and are among the highest reported for BLG on substrates.
Another indication of the high-quality of the sample can be deduced from 
 two-terminal quantum Hall measurements. In Fig.~\ref{fig01}(d) we show the transconductivity $d\sigma/dn$ as function of perpendicular $B$-field and carrier density $n=\alpha \Delta V_g$. Dashed lines are marking the filling factors $\nu = \pm12, \pm8, \pm4$. The eightfold degenerate zero Landau level unambiguously confirms the bilayer nature of the investigated  flake~\cite{nov06}. Additionally, we observe a degeneracy lifting into doubly degenerate Landau levels at $B \approx 2.5$~T, and full degeneracy breaking at $B \approx 6$~T (see squares and dots in Fig.~\ref{fig01}(d)), which 
is a direct signature of the high-quality of our sample and consistent with Refs. \cite{fel09,zha10}.
In Fig.~\ref{fig01}(e) we show the differential conductance $dI/dV_{b}$ as function of carrier density $n$ at constant magnetic field $B=2.5$~T (solid line).
The dashed line is the result of a numerical calculation for an ideal
BLG sample with the same aspect ratio of our device, following the approach of  Ref.~\cite{aba08} and using parameters discussed in~\cite{LLbroadening}. The experimental data closely follows the model of ideal BLG at charge carrier densities above $0.5 \times 10^{12}$~$\rm{cm}^{-2}$. 
The discrepancy at lower carrier density might be either explained by a small band gap resulting from an asymmetric doping of the top and bottom graphene layer or by the fact that symmetry breaking is not included in the model. 
From this calculation we also extract a gate lever arm of $\rm{\alpha} = 6.5 \times 10^{10}$ $\rm{cm}^{-2}$V$^{-1}$, which is in agreement with the slope of the dashed lines in Fig.~1(d) as well as with a plate capacitor model with the SiO$_2$ and the bottom hBN flake as gate dielectric.\newline
We next focus on weak localization (WL) measurements, from which we extract three fundamental time scales of our device: the phase coherence time $\tau_{\phi}$, the inter-valley scattering time $\tau_{i}$ as well as the intra-valley scattering time $\tau_*$. 
In Fig.~\ref{fig02} we show the experimentally observed WL dip, i.e. the change in conductivity 
at finite magnetic field with respect to the one at $B=0$~T:
 $\Delta\sigma(B)= \left\langle \sigma(n,B)-\sigma(n,B=0)\right\rangle_{\Delta n}$.
Fig. \ref{fig02}(a) shows $\Delta\sigma (B)$ at three different temperatures for a carrier density $n$ close to the charge neutrality point while Fig.~\ref{fig02}(b) shows similar data taken at $n = 3.2 \times 10^{11}~$cm$^{-2}$. Each trace is obtained by averaging 20 traces over a small density interval $\Delta n=2.6 \times 10^{9}~$cm$^{-2}$. The resulting magneto-conductivity traces 
exhibit an increase with positive and negative $B$-field followed by a saturation, which is consistent with previous WL experiments on BLG~\cite{gor07,che10,lia10}. 
To obtain the time scales mentioned above, we fit to
each trace the theoretical model for
WL in BLG (see dashed lines in Fig. 2) derived by
Kechedzhi \textit{et al.} \cite{kec07}:

\begin{eqnarray}
\small
\Delta\sigma(B) &=& \frac{e^2}{\pi h} \left[F\left(\!\frac{B}{B_{\phi}}\right)\!-\!F\!\left(\!\frac{B}{B_{\phi}\!+\!2B_i}\!\right)\!\right] \\ && + \frac{2e^2}{\pi h} F\!\left( \frac{B}{B_{\phi}\!+\!B_i+\!B_*}\! \right)\!. \nonumber
\label{eq01}
\end{eqnarray}

\begin{figure}[tb]\centering
\includegraphics[draft=false,keepaspectratio=true,clip,%
                   width=1\linewidth]%
                   {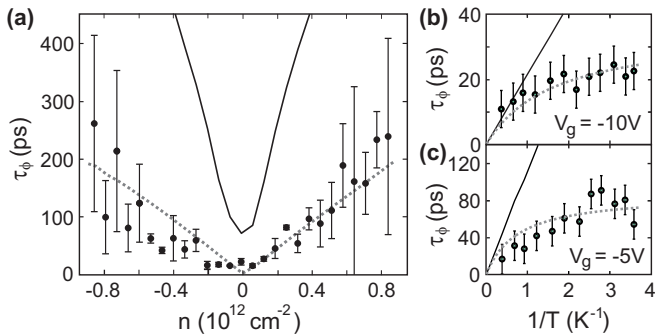}
\caption[fig03]{(a) shows the $n$ dependence of the phase coherence time $\tau_{\phi}$ at $T = 300$~mK. Here, the black line displays the theoretical result by AAK~\cite{alt82}. 
\newer{
The dashed line highlights the result modified due to a finite 
electron spin-flip scattering time $\tau_{sf}= \beta |n|$ with $\beta=3\times10^{-22}$~$\rm{cm}^2 \rm{s}$.} 
%
(b) and (c) display the temperature dependence of ${\tau}_{\phi}$ for carrier densities $n$ (b) very close to the charge neutrality point ($V_g$~=~-10~V) and (c) at $n = 3.2 \times 10^{11}$~$\rm{cm}^{-2}$ ($V_g$~=~-5~V). The solid and dashed lines resemble the same theoretical models as in panel~(a). From (c) we extract $\tau_{sf} = 88$~ps in good agreement with the fit in (a) from which we obtain $\tau_{sf} = 96$~ps for $n = 3.2 \times 10^{11}$~$\rm{cm}^{-2}$. All error bars are extracted from multiple measurements of each data point.
}
\label{fig03}
\end{figure}

Here $F(z)=\ln z+\Psi\left(\frac{1}{2}+\frac{1}{z} \right)$, where $\Psi$ is the digamma function, and $B_{\phi,i,*}$~=~$\hbar/ (4 e D \tau_{\phi,i,*})$.
The diffusion coefficient $D$ is given by the relation $D = \hbar g_{_{\square}}/ 4 m^*$, where $m^* = 0.033 \; m_0$ is the effective carrier mass in BLG~\cite{kos06}, and $g_{_{\square}}=\sigma h/e^2$ is the dimensionless conductivity at $B=0$~T.
We focus first on the behavior of the extracted phase-coherence time $\tau_{\phi}$. 
This time scale reaches a minimum of approx. 20~ps around the charge neutrality point (see Fig.~\ref{fig03}(a)). 
For larger carrier density, we observe 
a significant increase of $\tau_{\phi}$ by over one order of magnitude up to $\tau_{\phi}$~$\approx$~240~ps for $\left|n\right| = 0.8 \times 10^{12}~\rm{cm}^{-2}$.
Such a gate-tunable phase-coherence time - also measured on a second sample (see supplementary information) - has not yet been observed for single- and bilayer graphene. 
\newer{However, it is in qualitative agreement with a scattering mechanism based on electron-electron interactions as predicted by Altshuler, Aronov and Khmelnitsky (AAK) for a two-dimensional system, $\tau_{\phi}^{-1}= k_{B} T \ln g_{_{\square}} / (\hbar g_{_{\square}})$~\cite{alt82}, where $k_B$ is the  Boltzmann constant and $g_{_{\square}}$ is the dimensionless conductivity (introduced above), which can be directly extracted from the measured conductivity. Hence, without any adjustable parameter, we obtain the solid line in Fig.~\ref{fig03}(a).}
These values are a factor 3-4 larger than the values of $\tau_{\phi}$ extracted from our WL measurements,
 meaning that there must be an additional source for dephasing.
This becomes even more apparent when investigating the
temperature dependence of the extracted $\tau_{\phi}$ 
at different carrier densities, as shown in Figs.~3(b) and 3(c).
Similarly to Fig.~3(a), the solid line illustrates the estimates for 
\newer{$\tau_{\phi}$ obtained from
the AAK theory ($\tau_{\phi}^{-1} \sim T$)}. 
 Above $T$~=~400~mK (below $1/T~=~2.5$~$\rm{K}^{-1}$) \newer{the experimentally extracted} $\tau_{\phi}$
is inversely proportional to the temperature. However, at lower temperatures $\tau_{\phi}$ shows a saturation behavior.
\newer{To account for these discrepancies we follow Ref.~\cite{lara11} and include an additional inelastic electron spin-flip scattering time $\tau_{sf}$, leading to an overall scattering rate which is the sum of the spin-flip and the AAK decoherence rate, $\tau_{\phi}^{-1}~=~\tau_{sf}^{-1}~+~k_{B} T \ln g_{_{\square}} / (\hbar g_{_{\square}})$.}
We use this expression to estimate $\tau_{sf}$ and show
that by assuming a linear carrier density dependency ($\tau_{sf} \sim |n|$)
we obtain 
 good agreement with all our experimental data (see 
dashed lines in Figs. 3(a)-(c)). 
The extracted $\tau_{sf}$ values are in the order of 100~ps (see Fig.~S4 of the supplementary information), which is in agreement with earlier transport studies on graphene, where magnetic impurities were identified as a dominant phase scattering source~\cite{Lundeberg2013}. Although the apparent $n$-dependence of $\tau_{sf}$ is in contradiction to recent studies of spin scattering in single-layer graphene~\cite{jaro2014}, it may well be in agreement with resonant spin-flip scattering in BLG~\cite{privateJaro}.

\begin{figure}[b]\centering
\includegraphics[draft=false,keepaspectratio=true,clip,%
                   width=1\linewidth]%
                   {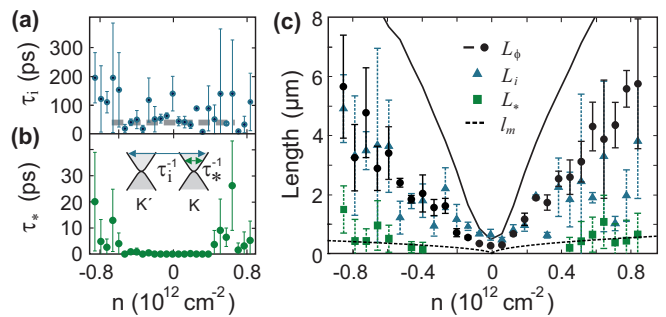}
\caption[fig04]{(a) and (b) show the $n$ dependence of the inter-valley scattering time $\tau_{i}$ and intra-valley scattering time $\tau_{*}$, respectively. (c) illustrates the corresponding length scales $L_{\phi}$, $L_{i}$, $L_{*}$ and the the mean free path $l_m$ (dashed line). The result of AAK is highlighted by the black solid line.}
\label{fig04}
\end{figure}

We focus now on the inter- and intra-valley scattering times $\tau_i$ and $\tau_*$ (see inset in Fig.~4(b)), which are related to the scattering processes that limit the mobility of our device (see discussion below).
The inter-valley scattering time $\tau_i$ as function of $n$ is shown in Fig.~\ref{fig04}(a). 
In contrast to the phase-coherence time, $\tau_{i}$ shows no clear dependence on the carrier density over a wide range of $n$.  This behavior is roughly consistent with the current understanding of inter-valley scattering processes in BLG which are caused by short-range scattering centers, such as lattice defects or ad-atoms. These can account for the large momentum transfer that is needed to scatter an electron from one valley to the other, and give rise to a density-independent  inter-valley scattering time~\cite{fer11}. Approximating observed values of $\tau_{i}$ with their weighted arithmetic mean  $\tau_i \approx$~40~ps (see dashed line in Fig.~\ref{fig04}(a)), we can obtain an estimate for the density of resonant scatterers (impurities) in our sample of $n_i = m^*/(8 \hbar \tau_i)\approx9\times10^7~{\rm cm}^{-2}$~\cite{fer11,shortrangeSc}. Taking into account the device geometry, this gives a total number of short-range scatterers 
of $\approx$100 which is a considerably low number compared to the total number of carbon atoms (roughly $10^{10}$ for 16$\times$7$\mu m$) and considering that the scatterers are most likely located at the edges of the BLG structure.
The intra-valley scattering time $\tau_*$ 
 is shown in Fig.~\ref{fig04}(b).  At low carrier density ($|n| < 0.4 \times 10^{12}$~$\rm{cm}^{-2}$) $\tau_*$ is smaller than 0.1~ps, indicating that in this regime the second term on the right hand side of Eq.~(1) is negligible and does not contribute to the overall shape of the WL dip, in agreement with previous findings~\cite{gor07}. 
At higher $n$, we obtain values of  $\tau_*$ = 5-30~ps, which are at least one order of magnitude smaller than $\tau_{\phi}$ and $\tau_{i}$.  Here, due to the strong trigonal warping effect in BLG, the intra-valley scattering time $\tau_*$  as extracted from WL measurements is not directly related to a chirality-breaking scattering process but is given by $\tau_{*}^{-1}=\tau_{w}^{-1}$+$\tau_{z}^{-1}$, where $\tau_{w}$ is the dominating trigonal-warping time and $\tau_{z}$ is the single-valley chirality-breaking time~\cite{kec07}.

More insight on the role of inter- and intra-valley scattering processes as mobility-limiting factors  can be obtained by looking at the corresponding characteristic length scales.  In Fig.~4(c) we plot $L_{\phi}$, $L_i$ and $L_*$, which are related to 
the respective time scale by $L_{\phi,i,*}$=$\sqrt{D\tau_{\phi,i,*}}$ together with the mean free-path, $l_m = \hbar \mu\sqrt{\pi n}/e$ (dashed line).  The phase-coherence length $L_{\phi}$ can be tuned up to 6~$\mu$m at sufficiently large densities, which, to the best of our knowledge, is significantly larger than all values previously reported in literature \cite{gor07,che10,lia10}. Most importantly, $L_{\phi}$ can be on the order of the sample width, making the presented BLG-hBN sandwich system interesting for future phase-coherent interference experiments. As for  the phase-coherence time, the experimental values of $L_{\phi}$ are reasonably close to the upper bound for the dephasing length set by the electron-electron interaction according to AAK theory (solid line in Fig.~\ref{fig04}(c)), but limited by the spin-flip scattering length
$L_{sf}$=$\sqrt{D\tau_{sf}}$, which is on the same order.

The inter-valley scattering length, $L_i$ in our sample (triangles in Fig.~4(c)) is about 0.4~$\mu$m at low carrier densities (roughly a factor 2 larger than $L_{\phi}$), and it increases up to 5~$\mu$m for larger $n$. In this regime $L_i$ exceeds $l_m$ by roughly one order of magnitude, ruling out inter-valley scattering 
 as the mechanism limiting mobility in BLG. 
 This observation clearly points at intra-valley scattering (being the only alternative) 
 as the mobility limiting process in high-quality BLG samples. Additionally, we find that the intra-valley scattering length $L_{*}$ (which is determined by both the trigonal warping effect and the single-valley chirality breaking time) is roughly one order of magnitude lower than $L_i$, but similar to $l_m$.
 
In literature, the main sources of intra-valley scattering have been associated with long-range disorder due to either charged impurities (Coulomb scatterers)~\cite{ada07,and06,nom07}, or  to local strain fluctuations~\cite{kat08}. However, differently from single-layer graphene and in agreement with earlier experiments on BLG samples \cite{xia10}, we can exclude that the limitations to mobility come from Coulomb scatterers. This is a consequence from the simple fact that, if Coulomb scattering was the limiting mechanism, the conductivity would show a different dependence on the carrier density  ($\sigma \sim n^{\alpha}$ and $1 < \alpha < 2$, where $\alpha$ is a density-dependent exponent~\cite{das10,nodas}), than the linear behavior $\sigma \sim n$ reported in Fig. 1(c) and other measurements~\cite{xia10,mor08}. 
Vice versa, it can be shown that local strain fluctuations in BLG lead to the correct dependance $\sigma \sim n$ \cite{paco}.  
 We therefore conclude that the electron mobility in our sample is  limited by intra-valley scattering events that are most likely caused by local strain fluctuations.  This conclusion agrees with evidence we have from Raman experiments that high-mobility samples exhibit reduced strain fluctuations~\cite{neu13}, as well as with recent studies on single-layer graphene, which also identified mechanical deformations as the main source of limiting mobility~\cite{nun14}. This in turn strongly suggests that the transport properties of both single- and bilayer graphene are hindered by the same physical mechanism. \newer{
To our knowledge, random strain fluctuations are most likely introduced by sample fabrication, where mechanical exfoliation, pre-deposition on rough SiO$_2$, graphene-substrate interactions and different thermal expansion coefficients may give rise to local mechanical deformations.
However, further work is needed to understand the detailed mechanisms.}

In conclusion we performed transport measurements on high-mobility bilayer graphene encapsulated in hBN. From WL measurements, we extract information on the dephasing time as well as the inter- and intra-valley scattering times and the corresponding characteristic length-scales. We observe phase-coherence lengths comparable with the sample size as well as phase-coherence times close to the values imposed by electron-electron interaction but limited by spin-flip scattering at low temperatures. Surprisingly, this spin-flip scattering time is more than an order of magnitude lower than observed in non-local spin-value measurements~\cite{spinvalve}.  Moreover, we can unambiguously conclude that  intra-valley scattering rather than inter-valley scattering is the limiting mechanism for electron transport in BLG, and we discuss strain fluctuations as the most probable source of mobility-limiting scattering processes.  

{Acknowledgments ---}
We thank R. V. Gorbachev and M. Goldsche for helpful discussions on the fabrication process, N. Freitag for help on the measurement system, J. Fabian, V. Fal'ko, F. Guinea, T. Ihn, S. V. Rotkin for fruitful discussions and F.~Haupt for helpful input on the manuscript. Support by the HNF, JARA Seed Fund, the DFG (SPP-1459), the ERC (GA-Nr. 280140) and the EU project Graphene Flagship (contract no. NECT-ICT-604391), are gratefully acknowledged.


\begin{thebibliography}{99}
\bibitem{LLbroadening}
Here, we used an aspect ratio of $\xi=L/W$~$=$~2.3 and the Landau level broadening $\lambda$ is chosen to be $\lambda$~=~0.5. 
Note, that $\lambda=0.5$ corresponds to a FWHM of the Shubnikov-de Haas oscillation $\Delta \nu$=2.35 on a filling factor scale. With $\nu=hn/eB$, $B$=2.5T the FWHM is $\Delta n$=1.42$\times 10^{11}$~$\rm{cm}^{-2}$ in carrier density which corresponds to $\Delta E$=5.1~meV.
\bibitem{shortrangeSc}
We extract $n_i$ from $\sigma_{dc}^{strong} = 4e^2/h \times k_F^2/(16 n_i)$~\cite{fer11} and applying the Drude model $\sigma=ne \mu$.
\end{thebibliography}
\end{document}